\documentclass{cimento}
\usepackage{natbib}
\usepackage[dvips]{graphicx}

\title{A simulation of solar convection at supergranulation scale}

\author{
M. Rieutord\from{laomp}\from{tarbes}\from{iuf},
H.-G. Ludwig\from{ll}\from{cral},
 T. Roudier\from{tarbes},
\AA. Nordlund\from{copen},
\atque R. Stein\from{chic}}

\instlist{\inst{laomp}
Observatoire Midi-Pyr\'en\'ees, 14 avenue E. Belin, 31400 Toulouse,
\inst{tarbes}Observatoire Midi-Pyr\'en\'ees, 57 avenue d'Azereix - BP 826 -
65008 Tarbes Cedex 
\inst{iuf} Institut Universitaire de France
\inst{ll}Lund Observatory, Box 43, 22100 Lund, Sweden
\inst{cral}C.R.A.L, \'Ecole Normale Sup\'erieure, 69364 Lyon, France
\inst{copen}Theoretical Astrophysics Center and Astronomical
Observatory/NBIfAFG, Juliane Maries Vej 30, DK-2100 Copenhagen, Denmark
\inst{chic}Michigan State university, East Lansing, MI 48824, USA
}

\PACSes{\PACSit{96.60.Mz}{Photosphere, granulation}
\PACSit{92.60}{Convection, turbulence, diffusion}}

\newcommand{\SG}{supergranulation}
                                                                          
\begin{document}

\maketitle

\begin{abstract}
We present here numerical simulations of surface solar convection
which cover a box of 30$\times$30$\times$3.2~Mm$^3$ with a resolution
of 315$\times$315$\times$82, which is used to investigate the dynamics
of scales larger than granulation.  No structure
resembling supergranulation is present; possibly higher Reynolds
numbers (i.e. higher numerical resolution), or magnetic fields, 
or greater depth are
necessary. The results also show interesting aspects of granular
dynamics which are briefly presented, like extensive p-mode ridges in
the k-$\omega$ diagram and a ringlike distribution of horizontal 
vorticity around granules.  At large scales, the horizontal velocity 
is much larger than the vertical velocity and the vertical motion is 
dominated by p-mode oscillations.
\end{abstract}

\section{Introduction}

The classical mechanism invoked to explain the origin of \SG\ is the
latent heat of the conversion of He$^{2+}$ into He$^+$ at
$\sim10000$~km below the sun surface. In a fluid at rest, such a heat
release (or absorption) may indeed trigger an instability which can
develop into motions at scales comparable to the \SG\ scale. However,
this reasoning might be oversimplified since the solar plasma is highly
turbulent and the above instability may be suppressed by the high
diffusion of turbulent motions. It may well be that \SG\ is a surface
phenomenon which results from a large-scale instability of a collection
of granules \cite[]{RRMR00}. Such an instability is a consequence of nonlinear
interactions between granules; large-scale perturbations feel the
effect of granulation as diffusion, namely turbulent viscosity. But as
shown by \cite{GVF94}, this viscosity may be negative which thus
triggers a large-scale instability. In the same vein, the very existence
of mesogranulation as a distinct scale of convection has been questioned
several times \cite[]{SB97,RRMR00,HBBBKPHR00}, casting doubts on the
importance of the first ionization of Helium in convection.  Another 
possibility is that supergranulation reflects the larger size of convective 
upflows with increasing scale height at greater depths.

These remarks show that the dynamics of scales larger than the
granulation are still largely not understood while they certainly play
an important role in the solar magnetism as shown by the interaction
between supergranulation and network.

In order to progress on these questions, we performed a numerical
simulation of compressible convection with radiative transport which
has a horizontal extent as wide as computationally affordable in order
to accommodate as large as a supergranule.  
The domain is, however, rather shallow having an aspect ratio of
$\approx 10$. In the following we present some preliminary results of
the simulation with emphasis on the dynamics of scales larger
than granulation.

\section{The simulation}

We simulate convection on a range of horizontal scales using a
radiation-hydro\-dynamics code developed by two of the authors
(Nordlund and Stein). A description of the code can be found
in \cite{SN98}.  Briefly, the code solves the equations of
hydrodynamics (mass, momentum, and energy conservation) together with
the equation of radiative transfer for a compressible, viscous medium.
The viscosity employed is an artificial hyperviscosity which
stabilizes the numerical scheme; our simulation falls into the class
of LES (Large Eddy Simulation).  Our aim is to simulate flows on
granular and larger scales. To this end certain trade-offs have been
made between physical realism and computational demands: the radiative
transfer is treated in the grey approximation and the horizontal
resolution is a rather coarse 95\,km. These restrictions made it
affordable to study a large volume representing a part of the solar
surface layers. The computational domain covers $30
\times 30\,\mathrm{Mm}^2$ horizontally, and it extends vertically from about
$0.5\,\mathrm{Mm}$ above $\tau=1$ to a depth of $2.7\,\mathrm{Mm}$
below this level. The computational grid comprises $315 \times 315$
nodes in each horizontal plane and 82 nodes in the vertical direction.
The lateral boundaries of the computational domain are periodic. The
upper boundary allows the transmission of acoustic waves. The lower
boundary allows a free in- and outflow of mass with the constraint of
a vanishing net mass flux across the boundary, and constrains the
horizontal velocity in inflows according to $du_\mathrm{H}/dt =
-u_\mathrm{H}/t_\mathrm {relax}$, where $t_\mathrm{relax}$ is half the
minimum advection time across the bottom zone. The entropy of
inflowing material is prescribed; this entropy together with the
gravitational acceleration and chemical composition are the basic
control parameters of the simulation. Note, unlike a
standard stellar atmosphere code, the effective temperature of the
model is an outcome of the computation. It
can be adjusted by changing the entropy of the inflowing material.

\begin{figure}[htp]
\centerline{\includegraphics[width=7cm]{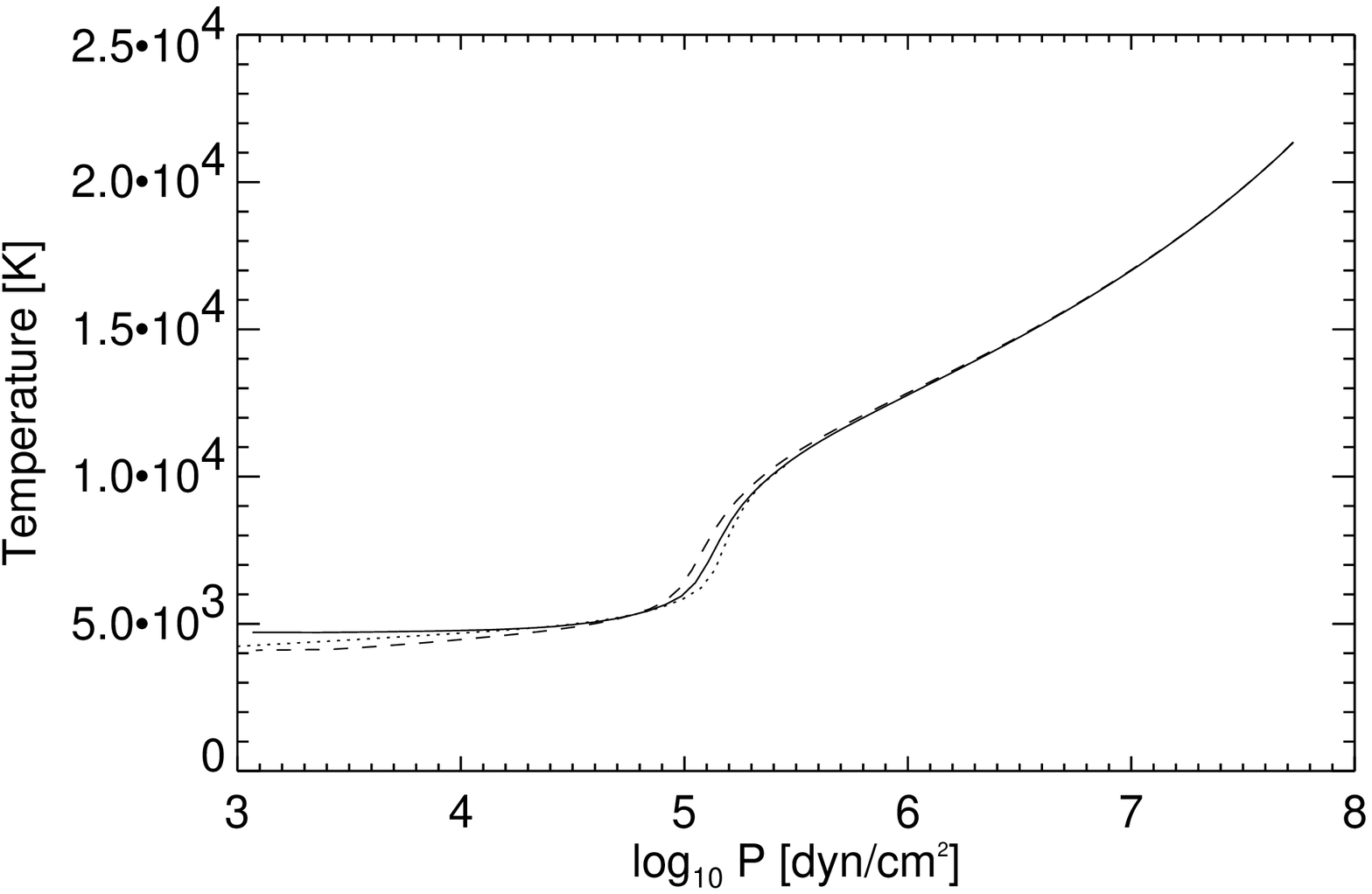}
\includegraphics[width=7cm]{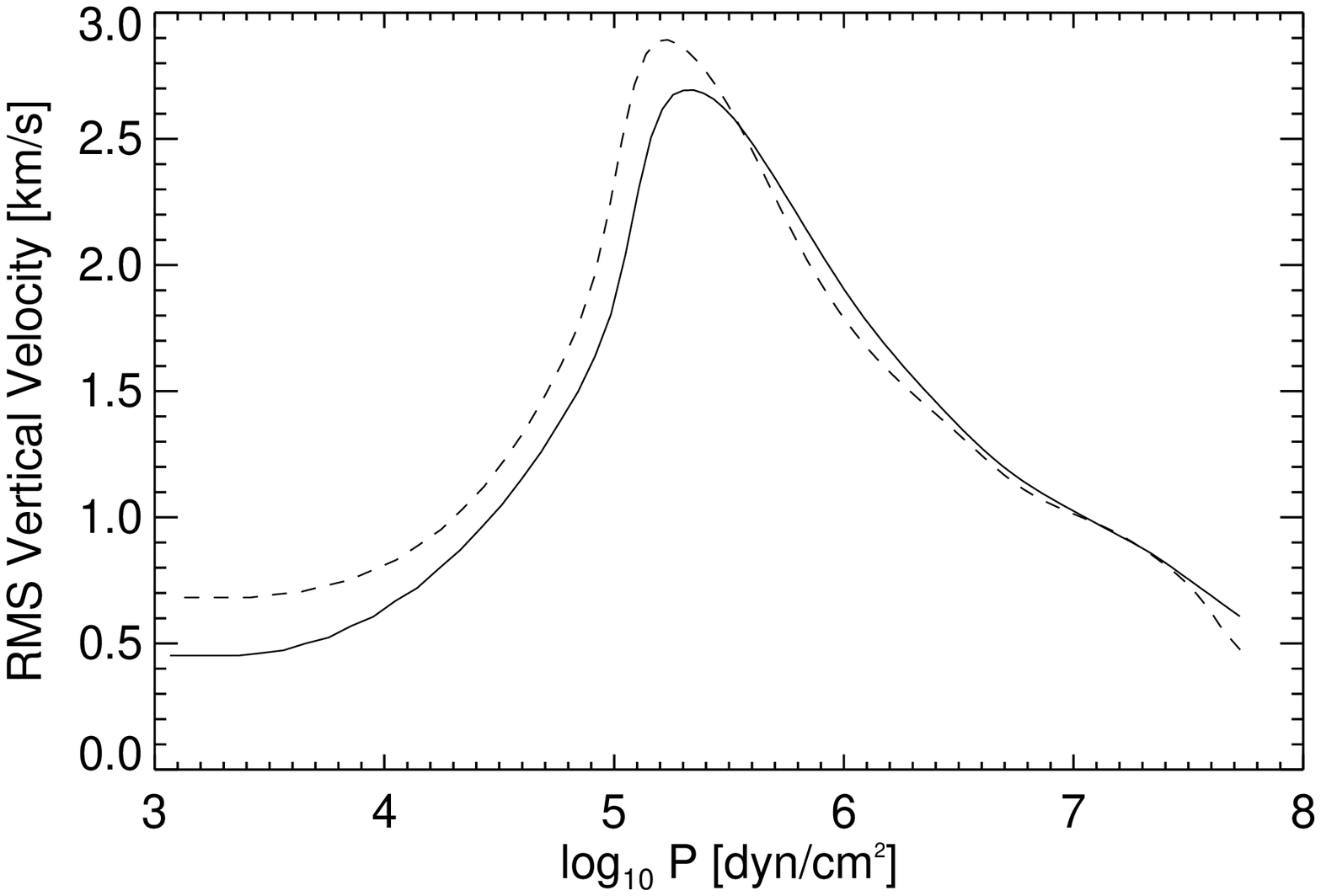}}
\caption[]{Profiles of temperature (left) and RMS vertical velocity
(right) issued from the simulation compared to profiles of a run with
twice the resolution (dashed line) and to an ATLAS9 model atmosphere
(dotted, temperature only).}
\label{profils}
\end{figure}

The radiative transfer is treated in strict LTE adopting grey
opacities which include contributions of spectral lines and are 
dependent on pressure and temperature. The non-locality of the
radiation field together with the geometry of the flow is taken into
account by solving the transfer equation at each time step employing a
modified Feautrier technique along a large number (here about 500,000)
of representative directions (``rays'', long characteristics)
traversing the computational domain. The equation of state takes into
account the ionization of hydrogen, helium, and other abundant
elements as well the formation of the $\mathrm{H}_2$
molecule. Opacities and the equation of state are derived from data of
the Uppsala stellar atmosphere package \cite{GBEN75}.

\begin{figure}
 \centerline{\includegraphics[width=7cm]{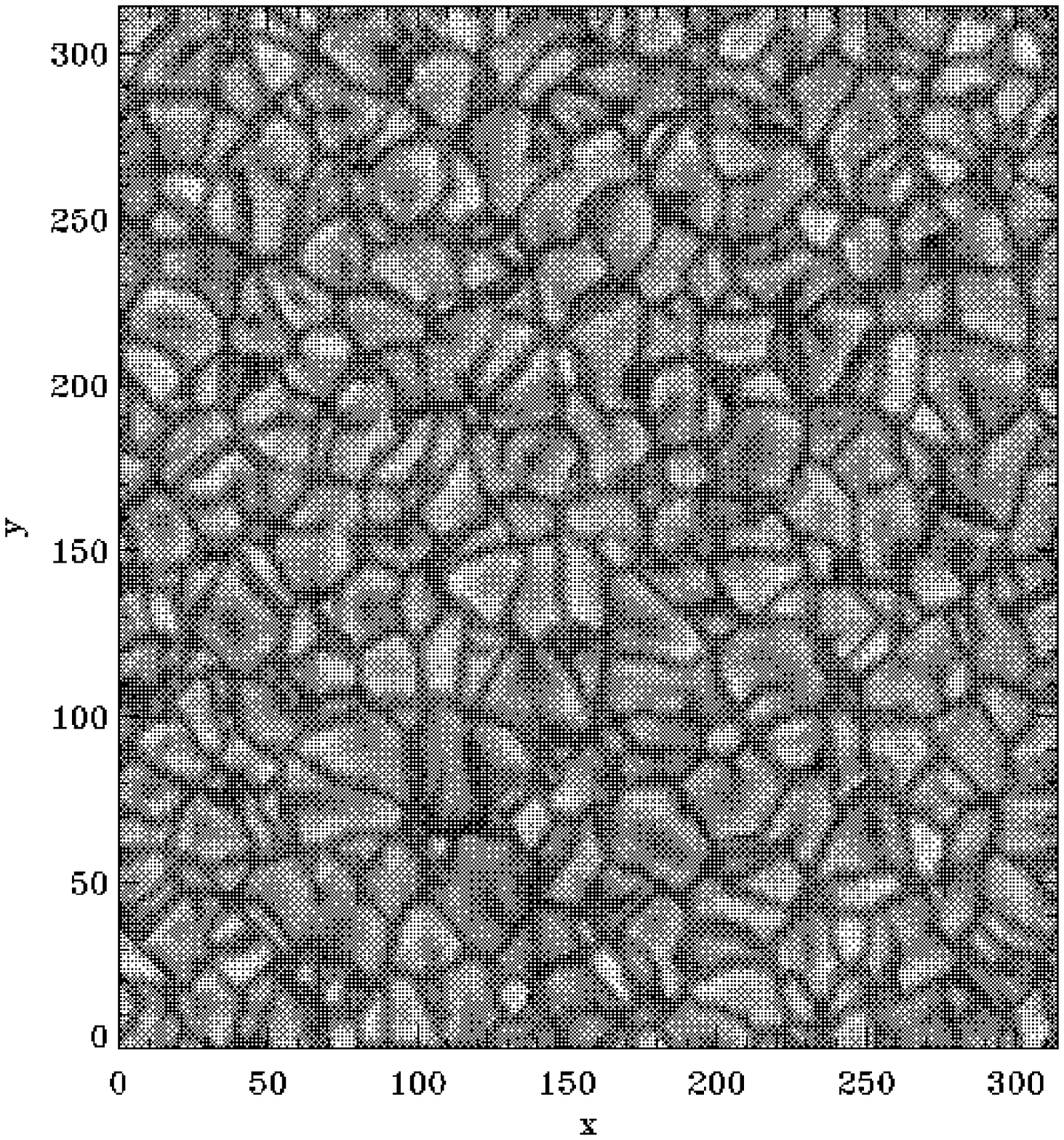}
 \includegraphics[width=7cm]{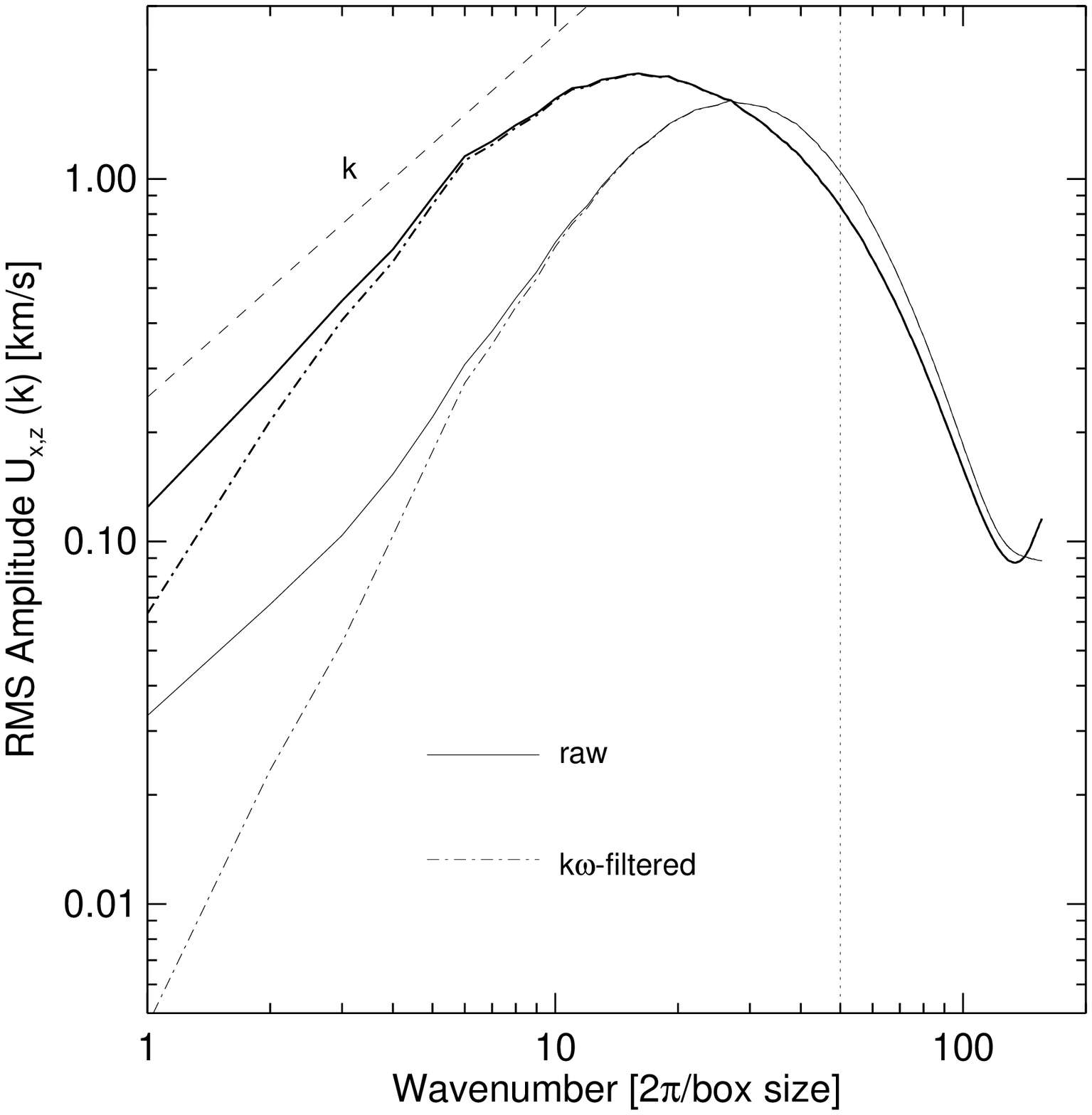}}
\caption[]{Left: Emerging intensity in the whole field. Right:
Time-averaged spatial spectra of the horizontal (thick lines) and
vertical (thin lines) velocity components at the $\tau=1$ level;
spectrum here designates $(kE(k))^{1/2}$ where $E(k)$ is the usual
spectral density of the squared velocity component. Spectra
including (solid) the p-mode oscillations, and after p-mode removal
(dashed-dotted) by sub-sonic filtering are shown. The dashed line
indicates a dependence $\propto k$, the dotted line the scale at which
numerical dissipation becomes important.}
\label{spectre}
\end{figure}

The large scale model was set up by tiling the plane with 5$\times$5
identical smaller scale models. Due to the periodic boundary conditions
this results in a flow field which is a valid solution of the governing
equations. The horizontal symmetry of the initial configuration was
broken by adding a small arbitrary velocity disturbance.  We simulated
almost 7\,hours of solar time from which about the last 5\,hours are
sufficiently evolved that the memory of the initial symmetry --- as
illustrated by figure~\ref{spectre} --- is lost and no longer plays a role
in the granular dynamics. This took 18,000\,CPU hours on a SGI Origin~2000
system (running 6 processors in parallel). The flow field was sampled
every 20\,s (solar time) producing a total of 110\,Gb of numerical data.

\section{Preliminary results}

In Fig.~\ref{profils} we compare the mean temperature and velocity
profile of the present simulation to one from a run with twice the
resolution but smaller horizontal extent. The temperature structure
shows some differences around $\log P=5.1$ (roughly corresponding to
unity optical depth) and the higher atmospheric layers. The differences
stem from the different treatment of the radiative transfer which is
grey in the present case as opposed to frequency-dependent in the
more resolved run. The ATLAS9 static model atmosphere shows
temperature deviations of similar order. Since we are here not aiming
at the best possible representation of the actual solar temperature
structure we consider the correspondence satisfactory.  The same holds
for the velocity profiles in Fig.~\ref{profils} which show some
increase of the velocity fluctuations with resolution, in particular
around the velocity maximum, an effect expected from previous
experience (see \cite{SN98}).

\begin{figure}[htp]
\centerline{\includegraphics[width=7cm]{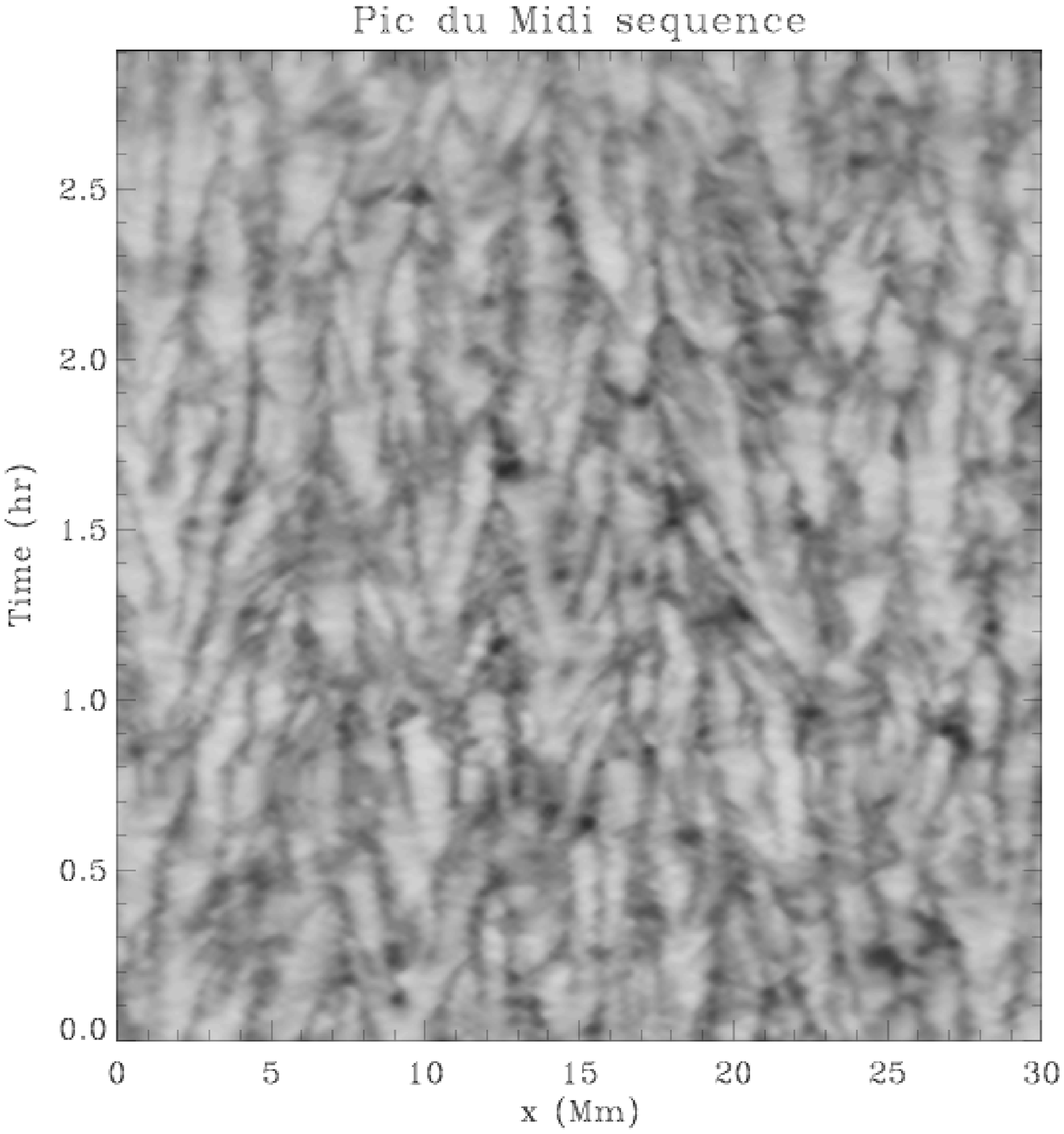}
\includegraphics[width=7cm]{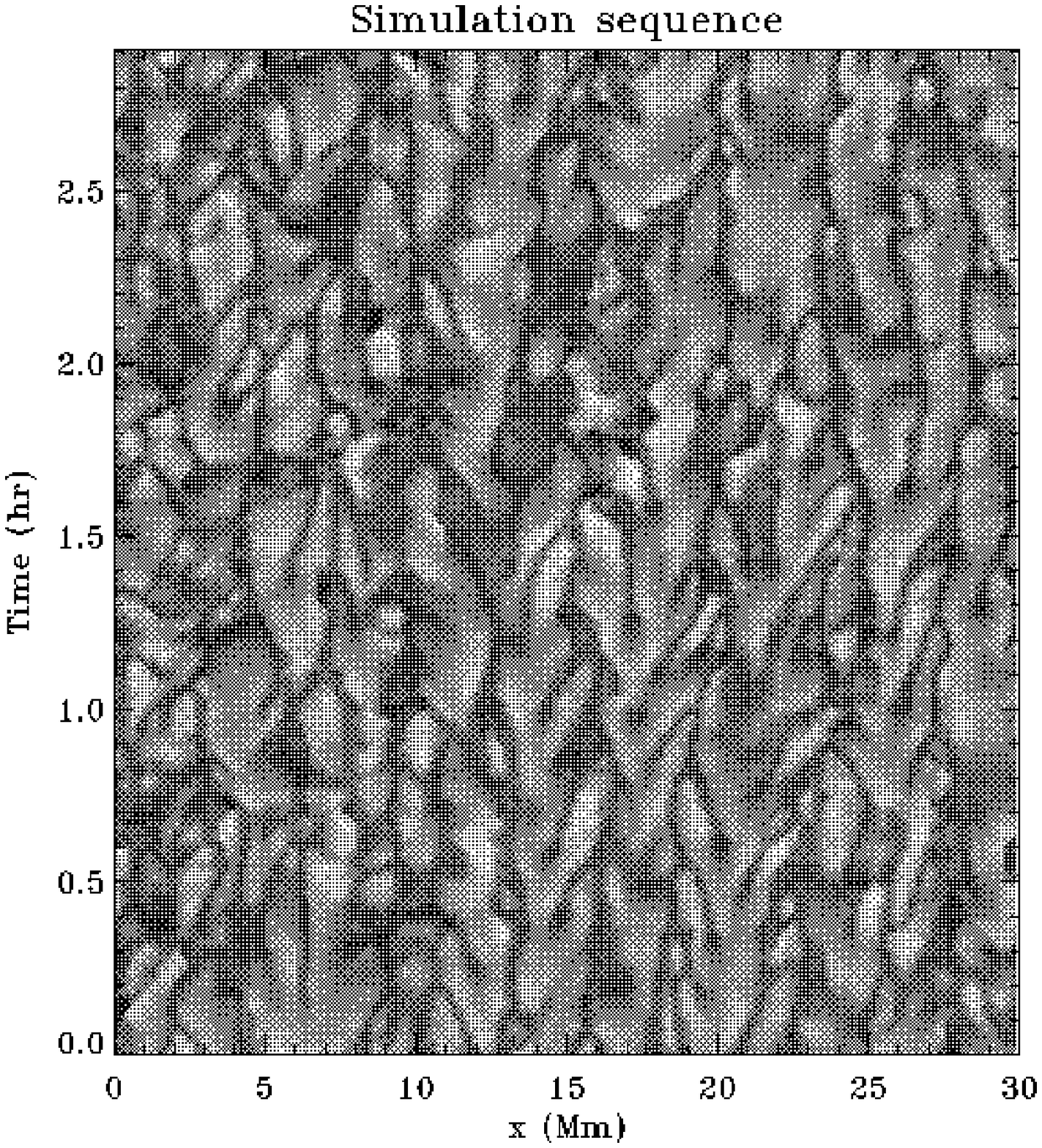}}
\caption[]{Grey scale diagram of intensity as a function of time and
space coordinates for an observed (left) and simulated (right)
sequence.  Both plots show that bright intensity features may last up
to 90~mn.}
\label{time_dist}
\end{figure}

In figure~\ref{spectre}, we plot the amplitude spectra 
($(kE(k))^{1/2}$, E(k) being the spectral density of velocity square)
of horizontal and vertical velocities near the $\tau=1$ level.  
Initially there was
no power present at scales larger than 6~Mm (k=1 -- 5). 
The time-averaged power spectrum shows three interesting features: 
First, there is no equivalent of
solar supergranulation.  This may be explained by many factors: the box
is too shallow, magnetic field is missing, resolution, i.e. the Reynolds
number, is too low or the length of the run is too short for
supergranulation to have developed.
Second, the horizontal velocities are mainly dominated by scales roughly
twice larger than those dominating vertical velocities.  This seems to
be related to the presence of strong downdrafts.  Third, the spectrum
for horizontal motion shows a break at 5Mm.  The origin of this break
needs further investigation.

\begin{figure}[htp]
\centerline{\includegraphics[width=7cm]{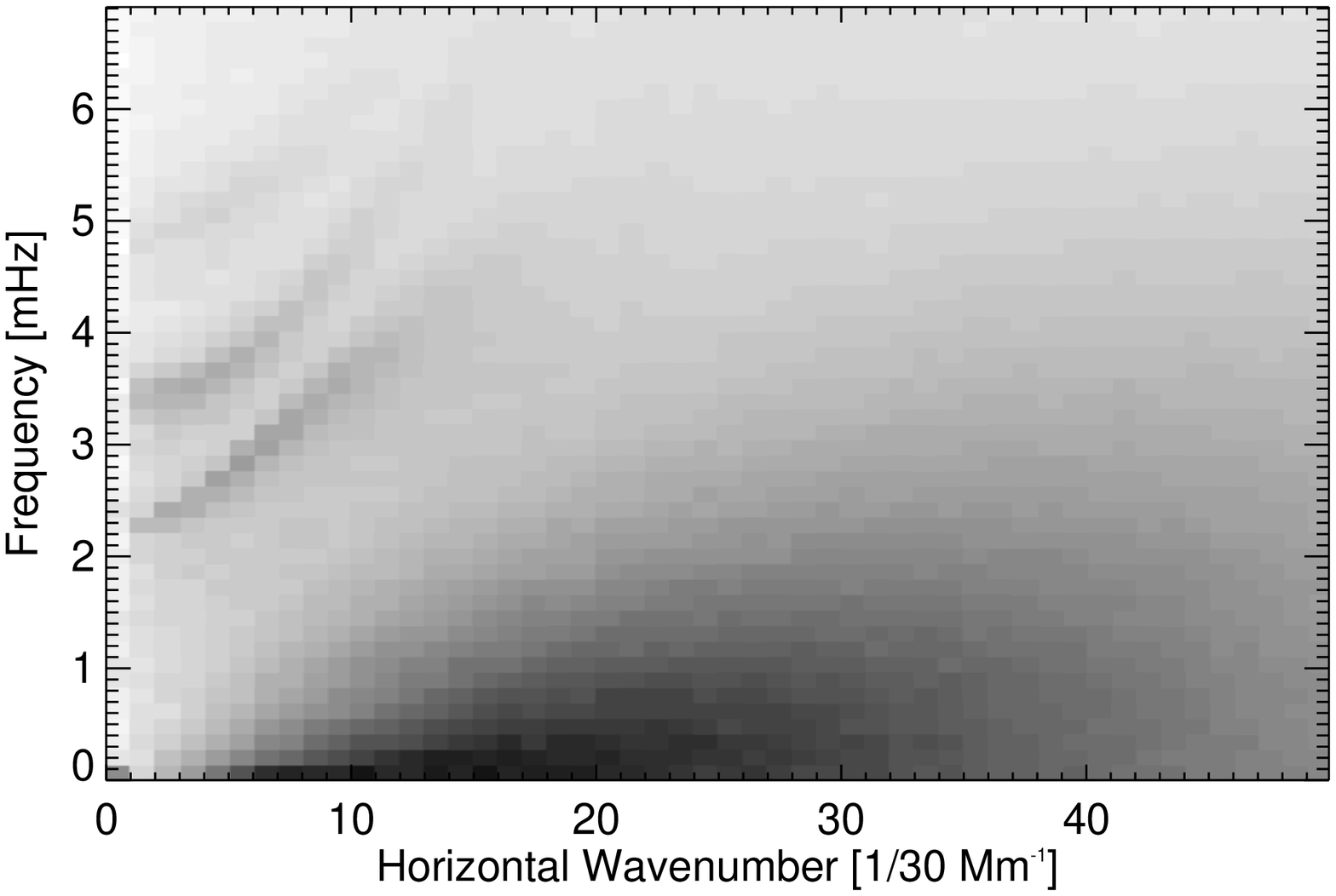}}
\caption[]{k--$\omega$ diagram of the vertical velocity around
$\tau=1$ showing the acoustic modes (dark ridges) excited by
convection (dark fuzzy cloud).}
\label{k_omeg}
\end{figure}

\begin{figure}[htp]
\centerline{\includegraphics[width=7cm]{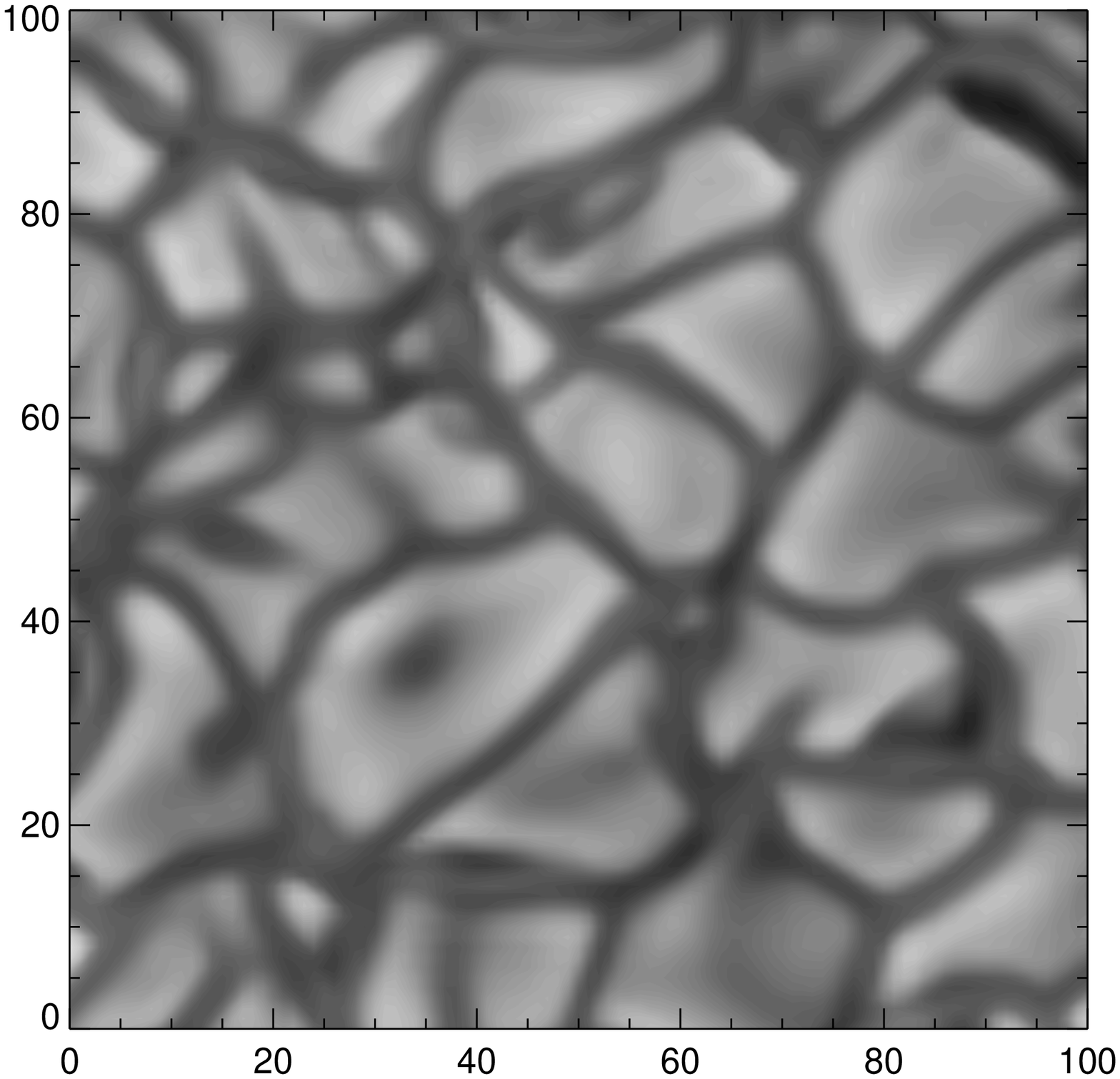}
\includegraphics[width=7cm]{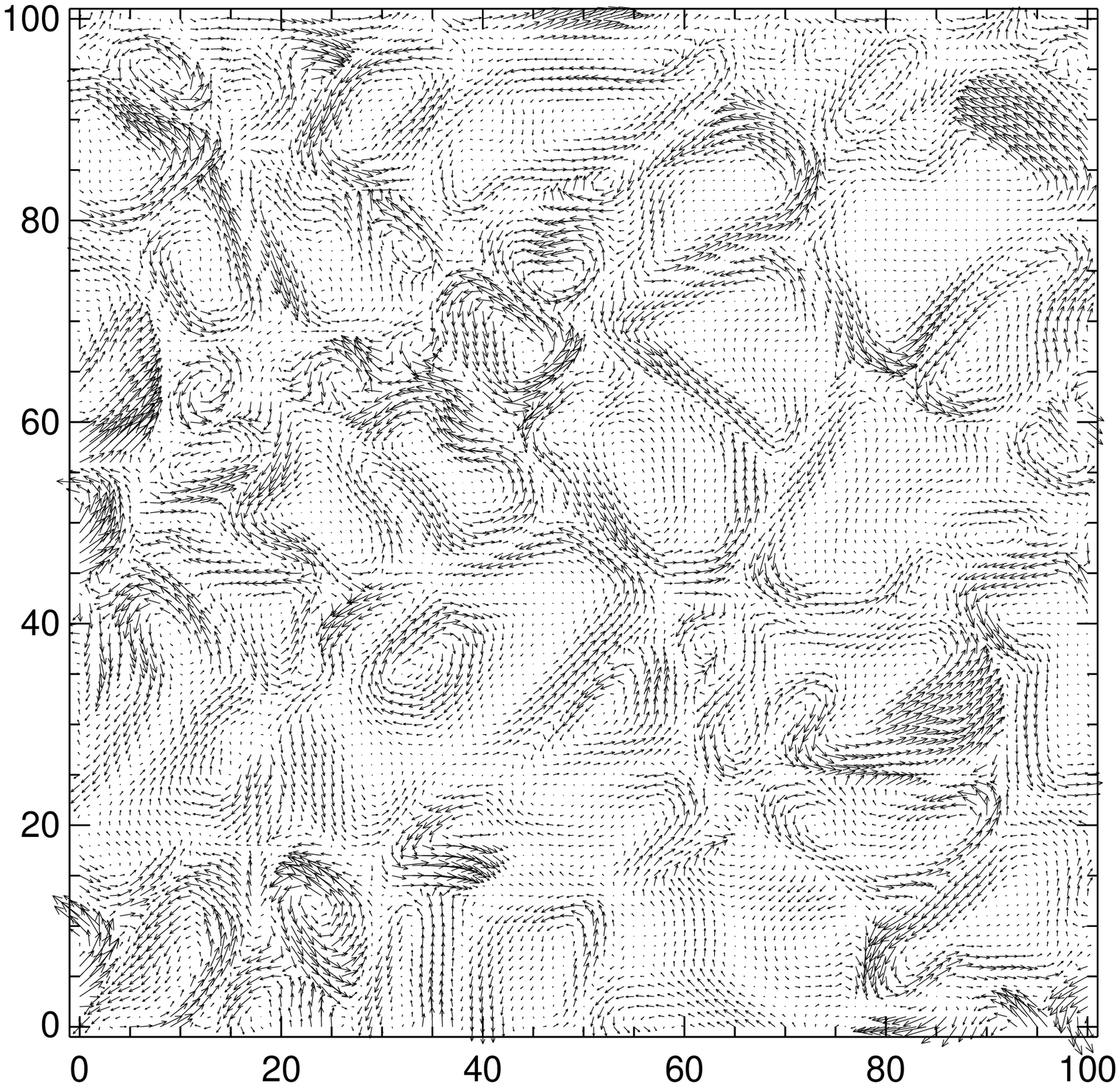}}
\caption[]{Intensity (left) and horizontal vorticity (right) in a data
subset of the simulation. x- and y-scale are pixels. We note that
granules are materialized by closed 'ropes' of vorticity which can be
approximated by vortex rings. Maximum arrow length corresponds to
0.076~s$^{-1}$, the RMS vorticity amounts to
0.011~s$^{-1}$ in the field of view. Note the exploding granule
located lower left off the center.}
\label{explod}
\end{figure}

Besides the question of the formation of large scale structures, the
simulation offers interesting ``side products''. For
instance, since the size of the domain is large enough to
contain many hundreds of granules, it was possible to test and assess
the relevance of granule tracking techniques \cite[see][]{NS88,
RRMV99} as a method for measuring plasma horizontal velocities at
scale larger than $\sim$3~Mm \cite[see][]{RRLNS01}.

In figure~\ref{time_dist} we computed a time-distance view of
granulation; such plots show that some granular features persist on
timescales of one hour or even more; such a behavior is also visible
in observational data and strengthens the confidence that the basic
dynamical properties of granulation are captured within the
simulation.  Another interesting dynamical feature is shown by the
k--$\omega$ diagram which reveals the acoustic modes excited by
convection in the box (see figure~\ref{k_omeg}). Despite its ``fuzzy''
appearance one should appreciate that it is perhaps the richest
k--$\omega$ diagram ever computed from a realistic convection
simulation including radiative transfer.

Finally, we plotted in figure~\ref{explod} a small field of view
centered on what we interpret to be an ``exploding granule'' (see 
Nordlund, 1985 for an explanation of ``exploding granules''); of
%Nordlund, Sol Phys. v100, p209, sect 2.7
course such an event is much smoother in the simulation than in the
real sun. We associate with this picture a plot of the horizontal
vorticity at the $\tau=1$ level (figure~\ref{explod}). 
This shows that granules are surrounded by a ring of vorticity where the
upflows turn over into the intergranular lanes.  This is the basis of
the vortex ring model of granules (\cite{parker92} and \cite{Are93b,Are94}).
Higher resolution simulations show that these vortex rings become 
turbulent and consist of tangled smaller vertex tubes \cite{SN98}. 
%This latter
%figure is interesting as it gives support to a simple model of
%granules, namely the vortex ring model, first proposed by
%\cite{parker92} and developed by \cite{Are93b,Are94}. Such a model is
%very much important for our subject as it gives a first step to
%elaborate a more realistic model of the flow at the sun's
%surface. Indeed, 
One can assemble such vortex rings and then analyse
the large-scale perturbations of such an assembly and determine
whether or not they are unstable.  Results along these lines of research
will be presented in future work.

\vspace*{-2mm}
\acknowledgments

Calculations were carried out on the CalMip machine of the
`Centre Interuniversitaire de Calcul de Toulouse' (CICT) which is
gratefully acknowledged. R.F.~Stein acknowledges financial support by
NASA grant NAG 5-9563 and NSF AST grant 98-19799.

\bibliography{../../biblio/bibnew}

\begin{thebibliography}{12}
\expandafter\ifx\csname natexlab\endcsname\relax\def\natexlab#1{#1}\fi

\bibitem[{Arendt(1993)}]{Are93b}
Arendt, S. 1993, Geophys. Astrophys. Fluid Dyn., 70, 161

\bibitem[{Arendt(1994)}]{Are94}
---. 1994, Astrophys. J., 422, 862

\bibitem[{Gama {et~al.}(1994)Gama, Vergassola, \& Frisch}]{GVF94}
Gama, S., Vergassola, M., \& Frisch, U. 1994, J. Fluid Mech., 260, 95

\bibitem[{{Gustafsson} {et~al.}(1975){Gustafsson}, {Bell}, {Eriksson}, \&
  {Nordlund}}]{GBEN75}
{Gustafsson}, B., {Bell}, R.~A., {Eriksson}, K., \& {Nordlund}, A. 1975,
  A. \& A., 42, 407

\bibitem[{Hathaway {et~al.}(2000)Hathaway, Beck, Bogart, Bachmann, Khtri,
  Petitto, Han, \& Raymond}]{HBBBKPHR00}
Hathaway, D., Beck, J., Bogart, R., Bachmann, K., Khtri, G., Petitto, J., Han,
  S., \& Raymond, J. 2000, Solar Phys., 193, 299

\bibitem[{November \& Simon(1988)}]{NS88}
November, L.~J. \& Simon, G.~W. 1988, Astrophys. J., 333, 427

\bibitem[{{Parker}(1992)}]{parker92}
{Parker}, E.~N. 1992, Astrophys. J., 390, 290

\bibitem[{Rieutord {et~al.}(2001)Rieutord, Roudier, Ludwig, Nordlund, \&
  Stein}]{RRLNS01}
Rieutord, M., Roudier, T., Ludwig, H.-G., Nordlund, {\AA}., \& Stein, R. 2001,
  A. \& A., 377, L14

\bibitem[{{Rieutord} {et~al.}(2000){Rieutord}, {Roudier}, {Malherbe}, \&
  {Rincon}}]{RRMR00}
{Rieutord}, M., {Roudier}, T., {Malherbe}, J.~M., \& {Rincon}, F. 2000, A.
  \& A., 357, 1063

\bibitem[{Roudier {et~al.}(1999)Roudier, Rieutord, Malherbe, \&
  Vigneau}]{RRMV99}
Roudier, T., Rieutord, M., Malherbe, J., \& Vigneau, J. 1999, A. \&
  A., 349, 301

\bibitem[{Stein \& Nordlund(1998)}]{SN98}
Stein, R.~F. \& Nordlund, {\AA}. 1998, Astrophys. J., 499, 914

\bibitem[{Straus \& Bonaccini(1997)}]{SB97}
Straus, T. \& Bonaccini, D. 1997, A. \& A., 324, 704

\end{thebibliography}

\end{document}